\documentclass[smallcondensed]{svjour3}

\usepackage{multirow}
\usepackage{amssymb}
\usepackage{amsmath, bbm}
\usepackage[figuresright]{rotating}
\usepackage{capt-of}

\newcommand{\La}{{\Lambda}}
\newcommand{\Si}{{\Sigma}}

\newcommand{\be}{\begin{eqnarray}}
\newcommand{\ee}{\end{eqnarray}}

\newlength{\feynwidth} 
\newlength{\feynwidthbig} 

\usepackage{xcolor} 
\usepackage{hyperref}
\usepackage{widetext}
\usepackage{cite}
\usepackage{graphicx}

\hypersetup{colorlinks=true,citecolor= blue}

\journalname{Few-Body Syst}

\usepackage[english]{babel}
\usepackage[autolanguage]{numprint}

\begin{document}

%\dochead{}

\title{Constraints on the $\Lambda$-neutron interaction 
from charge symmetry breaking in the \texorpdfstring{$\mathbf{^4_\Lambda \rm \bf He}$ - $\mathbf{^4_\Lambda \rm \bf H}$}{4-Lambda-He -- 4-Lambda-H} hypernuclei }

\author{Johann Haidenbauer \and  Ulf-G. Mei{\ss}ner \and Andreas Nogga}

\institute{Johann Haidenbauer \at 
IAS-4, IKP-3 and JHCP, Forschungszentrum J\"ulich, D-52428 J\"ulich, Germany \\ 
\email{j.haidenbauer@fz-juelich.de} 
\and 
Ulf-G. Mei{\ss}ner  \at HISKP and BCTP, Universit\"at Bonn, D-53115 Bonn, Germany \\
IAS-4, IKP-3 and JHCP, Forschungszentrum J\"ulich, D-52428 J\"ulich, Germany \\ 
Tbilisi State University, 0186 Tbilisi, Georgia \\
\email{meissner@hiskp.uni-bonn.de} 
\and 
Andreas Nogga \at 
IAS-4, IKP-3 and JHCP, Forschungszentrum J\"ulich, D-52428 J\"ulich, Germany \\ 
\email{a.nogga@fz-juelich.de}
}
%
%  Celebrating 30 years of the Steven Weinberg’s paper 
%  Nuclear Forces from Chiral Lagrangians
%
\date{draft: July 1, 2021}
\maketitle

%\begin{wideabstract}
\begin{abstract}
We utilize the experimentally known difference of the $\Lambda$ separation 
energies of the mirror hypernuclei ${^4_\Lambda \rm He}$ and ${^4_\Lambda \rm  H}$ 
to constrain the $\Lambda$-neutron interaction. 
We include the leading charge-symmetry 
breaking (CSB) interaction into our hyperon-nucleon interaction derived 
within chiral effective field theory at next-to-leading order. 
In particular, we determine the strength of the two arising CSB 
contact terms by a fit to the differences of the separation energies
of these hypernuclei in the $0^+$ and $1^+$ states, respectively. 
By construction, the resulting interaction describes 
all low energy hyperon-nucleon scattering data, the hypertriton 
and the CSB in ${^4_\Lambda \rm He}$-${^4_\Lambda \rm  H}$ accurately. 
This allows us to provide first predictions for the $\Lambda$n scattering 
lengths, based solely on available hypernuclear data. 
\end{abstract}
%\end{wideabstract}

%\eject
\section{Introduction} 
\label{sec:1} 
The large charge symmetry breaking (CSB), manifested in the 
differences of the $\La$-separation energies of the mirror
nuclei ${^4_\Lambda \rm He}$ and ${^4_\Lambda \rm H}$,
is one of the mysteries of hypernuclear physics. Already experimentally 
established in the early 1960s \cite{Dalitz:1964fu,Raymund:1964an}, for the 
ground ($0^+$) state, there is still no plausible and 
generally accepted explanation of it despite of numerous investigations 
\cite{Gibson:1973zza,Coon:1979si,Bodmer:1985km,Coon:1998jd,Nogga:2001ef,Gal:2015hh,Gazda:2015qyt,Gazda:2016ir}. 
Indeed, the separation-energy difference $\Delta E(0^+)$ of $340$~keV
\cite{Juric:1973zq}, benchmark for many decades, is about half of
the corresponding difference in the mirror nuclei ${^3\rm H}$ 
and ${^3\rm He}$ which amounts to $764$~keV \cite{Miller:2006tv}. 
However, while in the latter case about 90\,\% of the difference 
is due to the Coulomb force, its effect is rather small for the 
$A=4$ hypernuclei and, moreover, it goes into the wrong direction
\cite{Bodmer:1985km,Nogga:2001ef}. Thus, most of the CSB seen in the
$A=4$ hypernuclei must come from the strong interaction. 

The separation-energy difference for the excited $(1^+)$ state
was established with a measurement from 1979 \cite{Bedjidian:1979ih} 
and found to be $\Delta E(1^+) = 240$~keV. Thus, at that time, 
it looked as if CSB effects are practically spin (state) independent.
The situation changed considerably around 2015-2016 when new and
more refined data from experiments at J-PARC \cite{Yamamoto:2015avw}
and Mainz \cite{Esser:2015fm,Schulz:2016dt} became available. These 
led to the presently accepted values of 
$\Delta E(0^+) = 233\pm 92$~keV and 
$\Delta E(1^+) = -83\pm 94$~keV \cite{Gazda:2016ir}. 

As already indicated above, initial calculations of the 
${^4_\Lambda \rm He}$-${^4_\Lambda \rm H}$ binding energy difference
based on a two-body model \cite{Dalitz:1964fu,Coon:1979si} failed 
to describe the data. The principle CSB mechanism considered in
those studies consisted of $\La-\Si^0$ mixing. It facilitates
pion exchange between the $\La$ and the nucleons \cite{Dalitz:1964fu},
which is otherwise forbidden by isospin conservation. In addition,
contributions from $\eta-\pi^0$, $\omega-\rho^0$, etc., mixing
were taken into account. The situation did not improve with 
first more elaborate studies that employed four-body wave
functions from variational Monte Carlo calculations \cite{Coon:1998jd}.
And it remained also unchanged when the first full-fledged four-body
calculations based on the Faddeev-Yakubovsky approach became
available \cite{Nogga:2001ef}. The coupled-channel $\La N$-$\Si N$ 
interactions employed in the latter study, constructed by the 
Nijmegen group \cite{Maessen:1989hm,Rijken:1999fc}, all include 
$\La-\Si^0$ mixing as essential source of CSB. In addition, further 
sources of CSB such as the Coulomb interaction in the $NN$ and
$\Si N$ subsystems and the mass differences between $\Si^-$,
$\Si^0$ and $\Si^+$ were taken into account in Ref.~\cite{Nogga:2001ef}. 
But these calculations could only explain a fraction of the 
experimentally found CSB in $A=4$ hypernuclei. 
Very recently four-body calculations within the no-core shell model 
were presented by Gal and Gazda \cite{Gazda:2015qyt,Gazda:2016ir} which 
promised, finally, a solution to the CSB ``puzzle''. 
However, the CSB mechanism is somewhat unorthodox and rests on 
the assumption that the CSB part of the $\La N$ interaction can be
entirely and uniquely fixed by the $\La N\to \Si N$ transition 
potential \cite{Gal:2015hh}. 

In the present work, we study CSB in the hyperon-nucleon ($YN$) 
interaction within SU(3) chiral effective field theory (EFT)
\cite{Polinder:2006eq,Haidenbauer:2013oca,Haidenbauer:2019boi,Petschauer:2020urh},
which is an extension of Weinberg's idea suggested for nuclear forces
\cite{Weinberg:1990bf} to systems involving baryons with strangeness. 
In this approach, the long-range part of the interaction (due to exchange 
of pseudoscalar mesons) is fixed by chiral symmetry. The short-distance 
part is not resolved and effectively described by contact terms
whose strengths, encoded in low-energy constants (LECs), need 
to be determined by a fit to data \cite{Weinberg:1990bf,Epelbaum:2008ga,Machleidt:2011gh}. 
This notion applies to the charge-symmetry conserving as well as to 
the charge-symmetry breaking part of the interaction
\cite{Walzl:2000cx,Friar:2003yv,Epelbaum:2005ci}. 
Accordingly, in our investigation, we do not follow the aforementioned
procedure applied by Gal and Gazda. 
Rather, we fix the CSB part of the $\La N$ potential 
from the $A=4$ separation energies and then predict CSB effects 
for the elementary $\La p$ and $\La n$ interactions. 
We do not share the view of Gazda and Gal who consider this 
procedure basically as a tautology \cite{Gazda:2016ir}
but actually as an excellent tool to pin down the $\La n$ interaction, 
relying on and being consistent with available hypernuclear data.  

The paper is structured in the following way: in the subsequent
section, we give a detailed account of the CSB part of the $\La N$
interaction. Here we follow closely the arguments from analogous 
studies of the nucleon-nucleon ($NN$) system. 
Technical details of the treatment of the three- and four-body
systems are summarized in Sec. \ref{sec:fysummary}. 
In Sec. \ref{sec:results}, we explain how the CSB part is determined 
from the 
separation-energy differences in the $0^+$ and $1^+$ states of the 
$A=4$ hypernuclei. Specifically, considering those differences 
allows us to fix the low-energy constants of corresponding
CSB contact terms that arise at next-to-leading order in the
chiral expansion \cite{Epelbaum:2005ci}. Once those are 
established, predictions for $\La p$ and $\La n$ 
scattering lengths are presented. 
The paper ends with some concluding remarks. 

%%%%%%%%%%%%%%%%%%%%%%%%%%%%%%%%%%%%%%%%%%%%%

\section{Hyperon-nucleon interaction}
\label{sec:yninter}

\subsection{$YN$ interaction in chiral EFT}
\label{sec:ynnocsb}

For the present study, we utilize the $YN$ interactions from
Refs.~\cite{Haidenbauer:2013oca,Haidenbauer:2019boi}, derived within
SU(3) chiral EFT up to next-to-leading
order (NLO). At that order of the chiral expansion, the $YN$ potential 
consists of contributions from one- and two-pseudoscalar-meson
exchange diagrams (involving the Goldstone boson octet $\pi$, $\eta$, $K$)
and from four-baryon contact terms without and with two derivatives.
The two $YN$ interactions are the result of pursuing different 
strategies for fixing the low-energy constants (LECs) that determine 
the strength of the contact interactions.
In the $YN$ interaction from 2013 \cite{Haidenbauer:2013oca}, denoted 
by NLO13 in the following, all LECs have been fixed exclusively by a 
fit to the available $\Lambda N$ and $\Sigma N$ data. The other 
potential \cite{Haidenbauer:2019boi} (NLO19) has been guided 
by the objective to reduce the number of LECs that need to be fixed from 
the $Y N$ data by inferring some of them from the $NN$ sector via 
the underlying (though broken) SU(3) flavor symmetry. 
A thorough comparison of the two versions for a range of cutoffs 
can be found in Ref.~\cite{Haidenbauer:2019boi},
where one can see that the two $YN$ interactions yield essentially
equivalent results in the two-body sector. 
Note that there is no explicit CSB in the $\La N$ potential of 
the published $YN$ interactions. However, since the scattering amplitude 
in Refs.~\cite{Haidenbauer:2013oca,Haidenbauer:2019boi} 
is obtained from solving a coupled-channel Lippmann-Schwinger equation
in the particle basis, the mass differences between 
$\Si^-$, $\Si^0$, and $\Si^+$ enter and likewise the Coulomb
interaction in the $\Si^- p$ channel. Because of that isospin symmetry
is broken and the results for $\La p$ and $\La n$ scattering 
are (slightly) different. 

\subsection{CSB in chiral EFT}
\label{sec:csbeft}

\begin{figure}[t]
\begin{center}
\includegraphics[width=7.5cm]{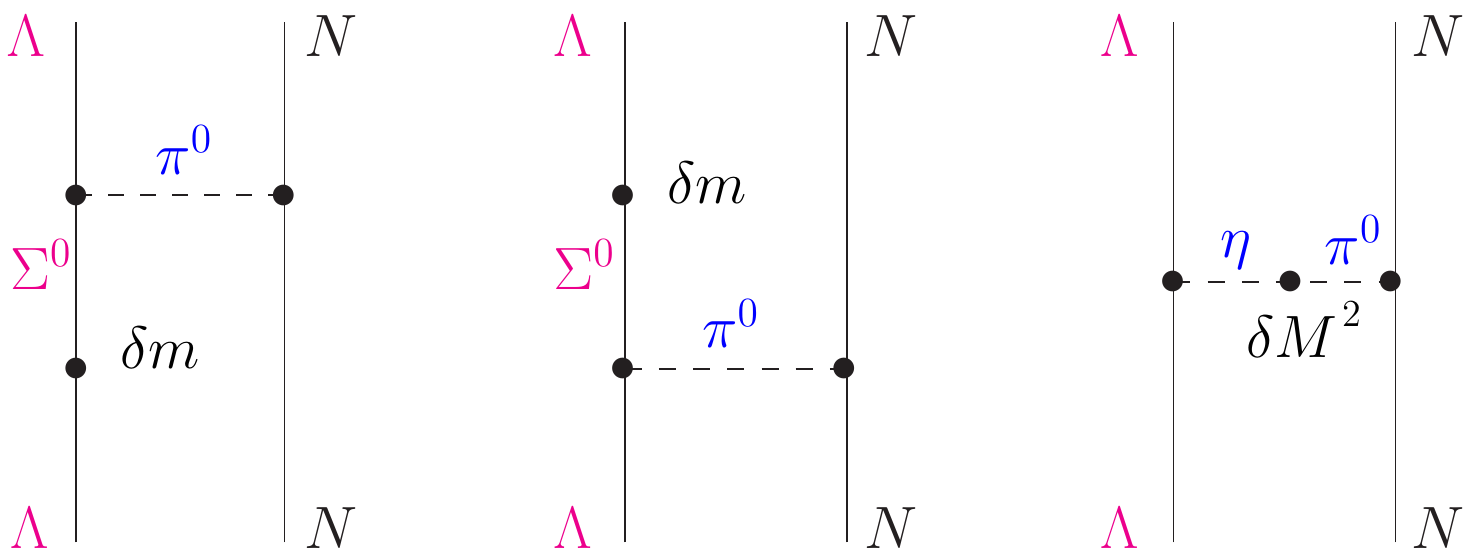}
\caption{CSB contributions involving pion exchange,
according to Dalitz and von Hippel \cite{Dalitz:1964fu},
due to $\La-\Si^0$ mixing (left two diagrams) and $\pi^0-\eta$ mixing
(right diagram).
}
\label{fig:csb1}
\end{center}
\end{figure}

As noted by Dalitz and von Hippel many decades ago~\cite{Dalitz:1964fu},
$\La - \Si^0$ mixing leads to a long-ranged CSB contribution to the $\La N$
interaction due to pion exchange, see Fig.~\ref{fig:csb1}.
The strength of the potential can be estimated from the electromagnetic mass matrices,
\begin{eqnarray}
\langle \Si^0|\delta m|\La\rangle & = & [m_{\Si^0}-m_{\Si^+}+m_p-m_n]/\sqrt{3}, \nonumber \\
\langle \pi^0|\delta M^2|\eta\rangle & = &  [M^2_{\pi^0}-M^2_{\pi^+}+M^2_{K^+}-M^2_{K^0}]/\sqrt{3} 
\end{eqnarray}
and subsumed in terms of an effective $\La\La\pi$ coupling constant
\begin{equation}
f_{\La\La\pi} = \left[
-2 \frac{\langle \Si^0|\delta m|\La\rangle}{m_{\Si^0}-m_{\La}}+ 
\frac{\langle \pi^0|\delta M^2|\eta\rangle}{M^2_{\eta}-M^2_{\pi^0}}
\right]\, f_{\La\Si\pi} \ .
\end{equation}
Based on the latest PDG mass values\cite{Zyla:2020zbs}, one obtains
\begin{equation}
f_{\La\La\pi}=f^{(\La - \Si^0)}_{\La\La\pi}+f^{(\eta-\pi^0)}_{\La\La\pi}
\approx (-0.0297 - 0.0106)\, f_{\La\Si\pi} \ . 
\label{Hippel}
\end{equation}
In this context, let us mention that there are also lattice 
QCD calculations of $\La-\Si^0$ mixing
\cite{Kordov:2019oer,Horsley:2014koa,Gal:2015iha,Horsley:2015lha}.

\begin{figure}[t]
\begin{center}
\includegraphics[width=4.5cm]{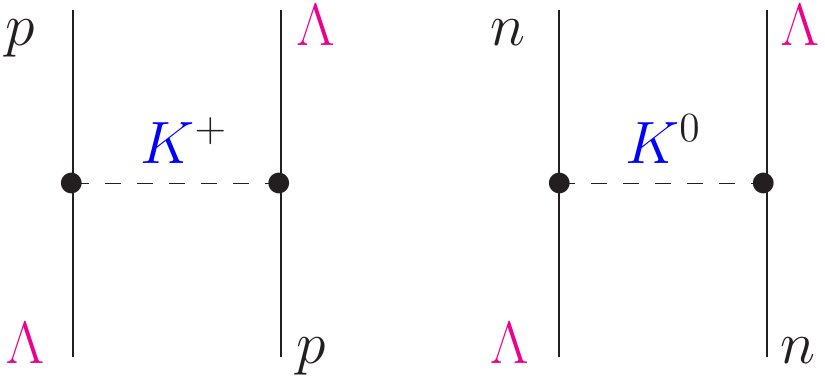}\qquad
\includegraphics[width=4.5cm]{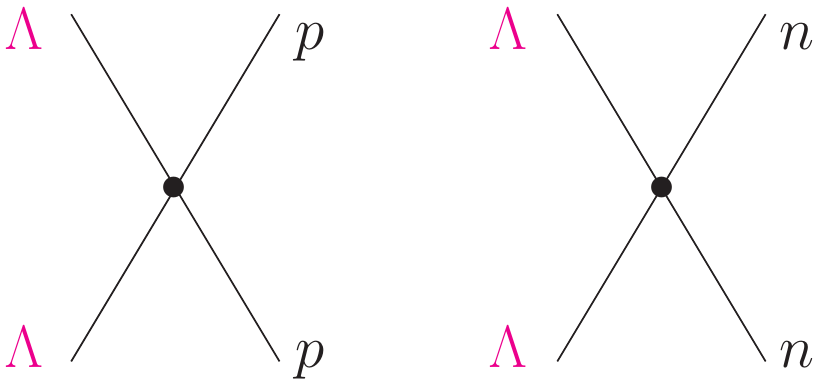}
\caption{CSB contributions from $K^{\pm}$/$K^0$ exchange
(left) and from contact terms (right).
}
\label{fig:csbct}
\end{center}
\end{figure}

In our implementation of CSB within chiral EFT, we follow closely the
arguments given in pertinent studies of isospin-breaking effects in 
the nucleon-nucleon ($NN$) system, see
Refs.~\cite{Walzl:2000cx,Friar:2003yv,Epelbaum:2005ci}.
According to Ref.~\cite{Friar:2003yv}, the CSB contributions at leading 
order are characterized by the parameter 
$\epsilon M^2_\pi / \Lambda^2 \sim 10^{-2}$, where
$\epsilon \equiv \frac{m_d-m_u}{m_d+m_u} \sim 0.3$ and 
$\Lambda\sim M_\rho$. In particular, one expects a potential strength 
of $V^{\rm CSB}_{BB} \sim (\epsilon M^2_\pi / \Lambda^2)\, V_{BB}$.
At order $n=2$ (NL{\O} in the notation of Ref.~\cite{Epelbaum:2005ci}),
there are contributions from isospin violation in the pion-baryon coupling constant,
which in the $\La N$ case arise from the aforementioned $\Sigma^0-\Lambda$ mixing as well as
from $\pi^0-\eta$ mixing. In addition,
there are contributions from short range forces (arising from $\rho^0-\omega$ mixing, etc.).
In chiral EFT, such forces are simply represented by contact terms involving LECs
(Fig.~\ref{fig:csbct} right) that need to be fixed by a  fit to data.
Contributions at $n=1$ (L\O) are due to a possible Coulomb interaction between the baryons
in question and due to mass differences between $M_{\pi^\pm}$ and $M_{\pi^0}$.
Such contributions do not arise in the $\Lambda N$ system. However, in the extension to
SU(3), there is CSB induced by the $M_{K^\pm}$-$M_{K^0}$ mass difference, see
left side of Fig.~\ref{fig:csbct}. 
We take that into account in our calculation, since it is formally
at leading order. But because the kaon mass is rather large compared to the mass
difference, its effect is actually very small.
For a general overview, we refer the reader to Table~I in Ref.~\cite{Epelbaum:2005ci}.

The CSB part of the $\La N$ potential at NL{\O} is given by
\begin{widetext}
\begin{eqnarray}
V^{CSB}_{\La N\to \La N}
&=& \Bigg[ -f^{(\La-\Si^0)}_{\La\La\pi} f_{NN\pi}\frac{\left(\mbox{\boldmath $\sigma$}_1\cdot{\bf q}\right)
\left(\mbox{\boldmath $\sigma$}_2\cdot{\bf q}\right)}{{\bf q}^2+M^2_{\pi^0}} \nonumber\\
& - & f^{(\eta-\pi^0)}_{\La\La\pi} f_{NN\pi}\left(\mbox{\boldmath $\sigma$}_1\cdot{\bf q}\right)
\left(\mbox{\boldmath $\sigma$}_2\cdot{\bf q}\right)
\left( \frac{1}{{\bf q}^2+M^2_{\pi^0}}-\frac{1}{{\bf q}^2+M^2_{\eta}} \right) \nonumber\\
&+&\frac{1}{4}(1-\mbox{\boldmath $\sigma$}_1\cdot\mbox{\boldmath $\sigma$}_2)\, C^{CSB}_{s}  
\,+\,\frac{1}{4}(3+\mbox{\boldmath $\sigma$}_1\cdot\mbox{\boldmath $\sigma$}_2)\, C^{CSB}_{t}  
\Bigg] \  \tau_N \ , 
\label{VCSB}
\end{eqnarray}
\end{widetext}
where $C^{CSB}_{s}$ and $C^{CSB}_{t}$ are charge-symmetry breaking
contact terms in the spin-singlet ($^1S_0$) and triplet ($^3S_1$) partial waves, 
respectively, and $\tau_p = 1$ and $\tau_n =-1$.
In the treatment of the contribution from $\pi^0-\eta$ mixing, we follow
Ref.~\cite{Coon:1979si}.

Besides the CSB in the $\La N$ potential, there is also some effect due to the
coupling to the $\Si N$ channel. Specifically, the coupled-channel Lippmann-Schwinger (LS)
equation is solved in the particle basis and the physical masses of the
$\Si$'s and $p$, $n$ are used. In addition, the Coulomb interaction in the
$\Si^-p$ channel that couples to $\La n$ is taken into account. As
mentioned above, these effects are already considered in our 
standard calculation \cite{Haidenbauer:2013oca,Haidenbauer:2019boi}
and lead to a small but noticeable CSB breaking in the $\La N$ results 
but also in case of ${^4_\Lambda \rm He}$ / ${^4_\Lambda \rm H}$ \cite{Haidenbauer:2007ra,Nogga:2013pwa,Nogga:2019bwo}.
CSB contributions to the $\La N$ potential from (irreducible) two-pion exchange,
which in our case not only involves the $p$-$n$ mass difference but also the one
for $\Si^+$-$\Si^0$-$\Si^-$, are expected to be small \cite{Friar:2003yv} and,
therefore, omitted in the present study. Note that the corresponding two-pion
exchange diagrams involve either $\pi^{\pm}$ or $\pi^0$ so that the
pion-mass difference does not enter.

We note already now that the value for $f_{\La\La\pi}$ given in Eq.~(\ref{Hippel}) 
as well
as the CSB LECs used in the actual calculation are in line with the aforementioned
order-of-magnitude estimate by Friar et al.~\cite{Friar:2003yv}.

%%%%%%%%%%%%%%%%%%%%%%%%%%%%%%%%%%%%%%%%%%%%%%%%%%%%%%%%%%%

\section{Faddeev-Yakubovsky equations}
\label{sec:fysummary} 

Our predictions for $A=3$ and $A=4$ systems 
are based on solutions of 
Faddeev-Yakubovsky equations in momentum space \cite{Nogga:2001ef, NoggaPhD:2001,Nogga:2013ey}. Here, we just briefly summarize
our way of solving the Yakubovsky equations for $A=4$ hypernuclei. 
For one hyperon and three identical nucleons, the Schrödinger equation 
can be rewritten in a set of five coupled Yakubovsky equations 
\begin{eqnarray}
\label{eq:yak}
  | \psi_{1A} \rangle & =  & G_0 t_{12} (P_{13}P_{23}+P_{12}P_{23}) \nonumber \\ 
  & &\qquad\qquad         \left[  | \psi_{1A}\rangle + | \psi_{1B}\rangle + | \psi_{2A}\rangle \right] \nonumber \\[5pt]  
    | \psi_{1B} \rangle & = & G_0 t_{12}  
           \left[ (1-P_{12}) (1-P_{23}) | \psi_{1C}\rangle \right. \nonumber \\[5pt] 
    && \qquad \qquad \left. + (P_{13}P_{23}+P_{12}P_{23}) | \psi_{2B}\rangle \right] \nonumber \\[5pt]  
    | \psi_{1C} \rangle & = & G_0 t_{14}  
           \left[ | \psi_{1A}\rangle + | \psi_{1B}\rangle + | \psi_{2A}\rangle              -P_{12} | \psi_{1C}\rangle \right. \nonumber \\[5pt] 
 && \qquad \qquad  \left. +P_{13}P_{23} | \psi_{1C}\rangle 
                   +P_{12}P_{23} | \psi_{2B}\rangle \right] \nonumber \\[5pt]  
  | \psi_{2A} \rangle & = & G_0 t_{12} 
           \left[ (P_{12}-1)P_{13}  | \psi_{1C}\rangle + | \psi_{2B}\rangle \right] \nonumber \\[5pt]    
  | \psi_{2B} \rangle & = & G_0 t_{34} \left[  | \psi_{1A}\rangle + | \psi_{1B}\rangle + | \psi_{2A}\rangle \right] \ .
\end{eqnarray}
For the solution of this problem, we distinguish the 
proton and neutron and the $\Sigma$ masses in 
the free propagator $G_0$ and for the solution of the Lippmann-Schwinger equations for the $t$-matrices $t_{ij}$ that are 
embedded into the four-baryon system. The reduction to only five Yakubovsky equations is possible because of the identity of the nucleons which allows to relate different Yakubovsky components to each other using permutation operators $P_{ij}$ that interchange the quantum numbers of nucleon $i$ and $j$.

The five remaining Yakobovsky components $| \psi_{1A}\rangle $, $| \psi_{1B}\rangle $, $| \psi_{1C}\rangle $, $| \psi_{2A}\rangle $ and $| \psi_{2B}\rangle $
are expanded in terms of their natural Jacobi coordinate
%\begin{widetext}
\begin{eqnarray}
\label{eq:jacmom}
    \textstyle  | p_{12} p_3 q_4 \alpha_A \rangle 
    & = & \textstyle \left| p_{12} p_3 q_4  \left[ \left[ (l_{12}s_{12})j_{12} \left(l_3 \frac{1}{2} \right) I_3 \right] j_{123}  \left(l_4 \frac{1}{2} \right) I_4 \right] J \left[ (t_{12} \frac{1}{2}) \tau_{123} t_Y \right] T M_T \right\rangle \ \nonumber \\[5pt]
    \textstyle  | p_{12} p_4 q_3 \alpha_B \rangle 
    & = & \textstyle  \left| p_{12} p_4 q_3  \left[ \left[ (l_{12}s_{12})j_{12}  \left(l_4 \frac{1}{2} \right) I_4 \right] j_{124}  \left(l_3 \frac{1}{2} \right) I_3 \right] J  \left[ (t_{12} t_Y ) \tau_{124} \frac{1}{2} \right] T M_T \right\rangle \ \nonumber \\[5pt]
    \textstyle  | p_{14} p_2 q_3 \alpha_C \rangle 
    & = & \textstyle  \left| p_{14} p_3 q_4  \left[ \left[ (l_{14}s_{14})j_{14}  \left(l_2 \frac{1}{2} \right) I_2 \right] j_{124}  \left(l_3 \frac{1}{2} \right) I_3 \right] J  \left[ (t_{14} \frac{1}{2}) \tau_{124} \frac{1}{2} \right] T M_T \right\rangle \ \nonumber \\[5pt]
    \textstyle  | p_{12} p_{34} q \beta_A \rangle 
    & = & \textstyle  \left| p_{12} p_{34} q  \left[ \left[ (l_{12}s_{12})j_{12}  \lambda \right] I \ (l_{34}s_{34})j_{34} \right] J \  (t_{12} t_{34}) T M_T 
        \right\rangle \ \nonumber \\[5pt]
    \textstyle  | p_{34} p_{12} q  \beta_B \rangle 
    & = & \textstyle  \left| p_{34} p_{12} q  \left[ \left[ (l_{34}s_{34})j_{34}  \lambda \right] I \ (l_{12}s_{12})j_{12} \right] J  \ (t_{34} t_{12}) T M_T 
        \right\rangle  \ .
\end{eqnarray}

%\end{widetext}
%
There are two types of Jacobi coordinates required. The 
first three basis sets are of the ``3+1'' type. Here, three momenta $p_{ij}$, $p_k$ 
and $q_l$ are required that are relative momenta within the pair $ij$, of particle 
$k$ with respect to pair $ij$ and of particle $l$ with respect to the 
three-body subsytem $ijk$. Corresponding orbital angular momenta $l_{ij}$, $l_j$ and $l_k$ are used to expand angular dependencies. $s_{ij}$ and $j_{ij}$ are 
the spin and total angular momentum of the two-body subsystem. 
We also introduce $j_{ijk}$ and $\tau_{ijk}$ for the total 
angular momentum and isospin of the three-body subsystem. $J$, $T$  and $M_T$ are the 
total angular momentum, isospin and third component of isospin of the 
four-baryon system. We have omitted the spins and isospins of the 
two baryons in the inner most subsystem since only $t_4=t_Y$ differs from 
${1}/{2}$. The last two basis sets are of the ``2+2'' type. 
Here, relative momenta of two two-body subsytems $p_{ij}$ and $p_{kl}$ 
are introduced together with angular momenta and isospins for these 
subsystems. Additionally, the relative momentum of the two pairs $q$ 
and its angular momentum $\lambda$ is required. In order to finally define 
the total four-body angular momentum in this case, an additional intermediate angular momentum $I$ is required and coupled to the other angular momenta as 
shown in Eq.~(\ref{eq:jacmom}).

Once the Yakubovsky components are found, we obtain the wave function by 
\begin{eqnarray}
  | \Psi \rangle & = & (1+P_{13}P_{23}+P_{12}P_{23}) | \psi_{1A} \rangle \nonumber \\[5pt]
        & & + (1+P_{13}P_{23}+P_{12}P_{23}) | \psi_{1B} \rangle \rangle \nonumber \\[5pt]
        & & + (1-P_{12}) (1+ P_{13}P_{23}+P_{12}P_{23}) | \psi_{1C} \rangle \nonumber \\[5pt]
        & & +(1+P_{13}P_{23}+P_{12}P_{23}) | \psi_{2A} \rangle\rangle \nonumber \\[5pt]
        & & +(1+P_{13}P_{23}+P_{12}P_{23}) | \psi_{2B} \rangle \ . 
\end{eqnarray}

For the solution of the Yakubovsky equations, the permutation operators $P_{ij}$
and transformations between different Jacobi coordinates Eq.~(\ref{eq:jacmom})
need to be evaluated. For these parts of the code, we use averages 
of the nucleon and $\Sigma$ masses. A comparison of the resulting 
energies and expectation values of the Hamiltonian shows that this approximation does not alter the results. 

For the numerical solution, the partial wave states have to be constrained. 
Here, we choose $j_{ij}\le 4$, $l_i\le 4$, $\lambda \le 4 $. Additionally, we restrict 
$l_{ij}+l_k+l_l \le 10$  and $l_{ij}+l_{kl}+\lambda \le 10$. 
We also only take the dominant isospin state $T=1/2$ into account. 

We checked carefully that, for the chiral interactions employed here, 
these constraints ensure that the numerical accuracy is better than 
10~keV for the energies entering the Yakubovsky equations (\ref{eq:yak}) and 20~keV for expectation values of the energy. It turns out that 
the additional isospin components with $T=3/2$ and $T=5/2$ induced by isospin breaking effects 
which we do not take into account here
lead to changes of the energy in this order and 
contribute most to this uncertainty.  

For the solution of this bound state problem, 
the Coulomb interaction in $YN$ and $NN$ is included. 
As discussed above, the Coulomb interactions only 
contribute a few keV to the separation energies. 
The same is true for the $n$-$p$
mass difference as has been 
shown in Ref.~\cite{Nogga:2002qp} 
for $^3$H-$^3$He. We therefore do not 
distinguish between contributions due to the $n$-$p$ mass difference and the one of the 
$\Sigma$'s in our results. In fact, the $n$-$p$ mass 
difference also contributes to the CSB of the core nucleus which we also do not separate from the other 
CSB contributions of the core.

\begin{table}
\renewcommand{\arraystretch}{1.4}
\caption{\label{tab:h3lamhe4lamh4lam} 
$^3_\Lambda$H, $^4_\Lambda$He and $^4_\Lambda$H
separation energies for NLO13 
and NLO19 for various cutoffs in combination with SMS N$^4$LO+ (450) \cite{Reinert:2018ip} and 
 different cutoffs for NLO13 and/or NLO19. No explicit CSB is included in the $YN$ 
 potentials. Energies are in MeV. 
}
\begin{tabular}{l|r|rr|rr}
\hline
      interaction &  $E_\Lambda(^3_\Lambda{\rm H})$ &  \multicolumn{2}{c|}{$E_\Lambda(^4_\Lambda{\rm He})$ } &  \multicolumn{2}{c}{$E_\Lambda(^4_\Lambda{\rm H})$ }\\
                     &     &   $J^\pi=0^+$ & $J^\pi=1^+$ & $J^\pi=0^+$ & $J^\pi=1^+$ \\                                      
\hline
NLO13(500)           & 0.13 &  1.71 &  0.80 &  1.66 & 0.78  \\
NLO13(550)           & 0.09 &  1.51 &  0.59 &  1.45 & 0.57   \\
NLO13(600)           & 0.09 &  1.48 &  0.59 &  1.43 & 0.56   \\
NLO13(650)           & 0.08 &  1.50 &  0.62 &  1.45 & 0.60   \\
\hline
NLO19(500)           & 0.10 &  1.65 &  1.23 &  1.63 & 1.23   \\
NLO19(550)           & 0.09 &  1.55 &  1.25 &  1.53 & 1.24   \\
NLO19(600)           & 0.10 &  1.47 &  1.06 &  1.44 & 1.05   \\
NLO19(650)           & 0.09 &  1.54 &  0.92 &  1.50 & 0.91   \\
\hline 
Expt.                & $0.13(5)$ \cite{Juric:1973zq} 
      & 2.39(3)\cite{Yamamoto:2015avw} & 0.98(3)\cite{Yamamoto:2015avw} 
      & 2.16(8)\cite{Schulz:2016dt} & 1.07(8) \cite{Schulz:2016dt} \\
\hline
\end{tabular}
\renewcommand{\arraystretch}{1.0}
\end{table}

As shown in previous calculations, the $\La$ separation energies
of light hypernuclei are only mildly dependent on the underlying
$NN$ interaction \cite{Nogga:2001ef,Haidenbauer:2019boi}.
Therefore, we employ in all of the calculations presented here the 
same chiral semi-local momentum-space-regularized (SMS)
$NN$ interaction of Ref.~\cite{Reinert:2018ip} at order N$^4$LO+
for a cutoff of $\La=450$~MeV. We have also used the Idaho interaction 
\cite{Entem:2003ft} for these calculations and have not found 
any significant changes of the CSB predictions. We note that  at LO, we find 
somewhat larger separation energies for $A=4$ hypernuclei for N$^4$LO+ 
compared to Idaho. Apparently, the missing repulsion at short distances 
at LO increases the sensitivity to configuration of the nucleons in the core. 
At NLO, the $NN$ force dependence of the separation energies is of the order 
of 100 keV and within the range expected from calculations based on 
phenomenological interactions \cite{Nogga:2001ef}.
To define a baseline, we summarize our results for the 
separation energies of $A=3$ and 4 hypernuclei for the 
original NLO13 and NLO19 interactions in 
Table~\ref{tab:h3lamhe4lamh4lam}.

Below we will present two kinds of results for the CSB. First, 
we will perform complete calculations for $^4_\Lambda$He and $^4_\Lambda$H.
This will allow us to obtain the CSB of the separation energies directly. 
But it does not allow for an easy separation of the different contributions 
to CSB. Second, we will use the wave function and Yakubovsky components 
for $^4_\Lambda$He of the original interactions 
to evaluate CSB perturbatively. For this, we calculate 
the expectation values of the differences of the kinetic energy and the $NN$  
and $YN$ potentials when $n$ and $p$ and $\Sigma^+$ and $\Sigma^-$ 
are interchanged. The total CSB of both calculations agrees to better 
than 10~\% for our standard calculations.

\section{Results}
\label{sec:results} 

As argued in Refs.~\cite{Haidenbauer:2019boi,Le:2019gjp}, the contribution 
of three-body forces (3BFs) is probably negligible for $A=3$, but
very likely becomes relevant for the more strongly bound $A=4$ system.
The dependence of the separation energies on the regulator (cutoff) is 
an effect of next-to-next-to-leading order (N$^2$LO) which includes 
also 3BFs \cite{Hammer:2012id}.
As can be seen from the results in Table~\ref{tab:h3lamhe4lamh4lam}, the variation is negligible for the 
hypertriton but can be as large as $200$--$300$~keV for $A=4$. 
Even larger is the difference between the two different realizations 
of $YN$ interactions: NLO19 and NLO13. For the $1^+$ state, the 
predictions of these two essentially equivalent realizations of the $YN$ 
interaction can differ by as much as $500$~keV. 
These variations in the predicted $A=4$ separation energies
have to be kept in mind and, ultimately, should be explained by  
similarly large 3BF contributions.

However, such 3BFs should have only a minor influence on the observed splittings
between the ${^4_\Lambda \rm He}$ and ${^4_\Lambda \rm H}$ states, 
which are primarily due to CSB two-baryon forces. 
This is the basic assumption in the strategy pursued below.  

\subsection{CSB in $A=4$ hypernuclei} 
\label{sec:csbexp}

To start with, we need to fix the CSB-breaking LECs
$C^{CSB}_{s}$ and $C^{CSB}_{t}$
in the $\La N$ interaction, cf. Eq.~(\ref{VCSB}). 
We do this by considering
the observed CSB splittings in the $A=4$ hypernuclei, 
defined in the usual way in terms of the separation energies 
\begin{eqnarray}
\Delta E(0^+) & = & E^{0^+}_{\La}({^4_\Lambda \rm He})
-E^{0^+}_{\La}({^4_\Lambda \rm H}), \nonumber \\
\Delta E(1^+) & = &  E^{1^+}_{\La}({^4_\Lambda \rm He})
-E^{1^+}_{\La}({^4_\Lambda \rm H}) \ . 
\end{eqnarray}

In our principal results, we aim at a reproduction of the present experimental
situation, based on the recent measurements of the 
${^4_\Lambda \rm H}$ $0^+$ state in Mainz \cite{Schulz:2016dt}
and the one of the ${^4_\Lambda \rm He}$ $1^+$-$0^+$ splitting at J-PARC \cite{Yamamoto:2015avw}, i.e.
$\Delta E(0^+) = 233\pm 92$~keV and $\Delta E(1^+) = -83\pm 94$~keV.
It is the same scenario as considered  by Gazda and Gal in Ref.~\cite{Gazda:2016ir}.
Below, we will refer to this choice as CSB1. 
In order to illustrate the effect of CSB in the $A=4$ hypernuclei on 
the underlying 
$\La N$ interaction, we consider also two other scenarios. One (CSB3)
corresponds to the situation after the publication of the J-PARC experiment \cite{Yamamoto:2015avw} but before the final results from Mainz became
available:  
$\Delta E(0^+) = 350\pm 50$~keV and $\Delta E(1^+) = 30\pm 50$~keV.
It is the status considered by Gazda and
Gal in Ref.~\cite{Gazda:2015qyt} and discussed in the review \cite{Gal:2016boi}.
In addition, we look at the situation up to 2014 (which will be labeled CSB2), namely
$\Delta E(0^+) = 350\pm 50$~keV and $\Delta E(1^+) = 240\pm 80$~keV
\cite{Bedjidian:1979ih}.
It is the one discussed by Gal in Ref.~\cite{Gal:2015hh} and, of course,
in all pre-2014 studies of CSB in the $A=4$ hypernuclei.
Note that the CSB splitting in the $1^+$ states in the scenarios CSB1 and 
CSB3 is compatible with zero, 
given the present experimental uncertainty.

\begin{table*}[tbp]
\caption{Comparison of different CSB scenarios, based on the $YN$
interactions NLO13 and NLO19 with cutoff $\Lambda = 600$~MeV. 
Results are shown for the original NLO interactions, with addition of 
OBE contribution to CSB, and for the scenarios CSB1, 
CSB2, CSB3 with added CSB contact terms. 
CSB1 corresponds to the present experimental status. 
Note that the $\chi^2$ for the NLO interactions differs slightly from the 
ones given in Refs.~\cite{Haidenbauer:2013oca,Haidenbauer:2019boi} because there the small
differences between $\La p$ and $\La n$ have not been taken into account. Small deviations of the CSB from values of the three
scenarios are due to using perturbation theory for fitting 
and using a smaller number of partial waves for fitting. 
}
\label{tab:ynprop}
\setlength{\tabcolsep}{1mm}
\renewcommand{\arraystretch}{1.4}
\centering \scriptsize
\begin{tabular}{|c|cc|cc|ccc|cc|}
\hline
&
$a^{\Lambda p}_{s}$ & $a^{\Lambda p}_{t}$ &
$a^{\Lambda n}_{s}$ & $a^{\Lambda n}_{t}$ &
$\chi^2(\Lambda p)$ & $\chi^2(\Sigma N)$ & $\chi^2({\rm total})$
& $\Delta E(0^+)$ & $\Delta E(1^+)$\\
\hline
\hline
NLO13  &-2.906 &-1.541 & -2.907 &-1.517 & 4.47 &12.34 & 16.81  &  58 &  24 \\
CSB-OBE&-2.881 &-1.547 & -2.933 &-1.513 & 4.39 &12.43 & 16.83  &  57 &  20 \\
CSB1   &-2.588 &-1.573 & -3.291 &-1.487 & 3.43 &12.38 & 15.81  & 256 & -53 \\
CSB2   &-3.983 &-1.281 & -2.814 &-0.948 & 4.51 &12.31 & 16.82  & 299 &  161 \\
CSB3   &-2.792 &-1.666 & -3.027 &-1.407 & 9.52 &12.41 & 21.93  & 370 &  56  \\
\hline
\hline
NLO19  &-2.906 &-1.423 & -2.907 &-1.409 & 3.58 &12.70 & 16.28 &  34 &  10 \\
CSB-OBE&-2.877 &-1.415 & -2.937 &-1.419 & 3.30 &13.01 & 16.31 &  -6 &  -7 \\
CSB1   &-2.632 &-1.473 & -3.227 &-1.362 & 3.45 &12.68 & 16.13 & 243 & -67 \\
CSB2   &-3.618 &-1.339 & -3.013 &-1.117 & 4.02 &12.09 & 16.12 & 218 &  129 \\ 
CSB3   &-2.758 &-1.546 & -3.066 &-1.300 & 7.49 &12.64 & 20.14 & 359 &  45 \\
\hline
\end{tabular}
\renewcommand{\arraystretch}{1.0}
\end{table*}

\begin{table*}[tbp]
\caption{Singlet ($s$) and triplet ($t$) $S$-wave scattering lengths
and $\chi^2$ values for the fits to the present experimental CSB splittings 
of $\Delta E(0^+)=233$~keV and $\Delta E(1^+)=-83$~keV (CSB1),
based on the $YN$ interactions NLO13 and NLO19. 
}
\vskip 0.2cm
\label{tab:A1}
\renewcommand{\arraystretch}{1.4}
\centering
\begin{tabular}{|c|cc|cc|ccc|}
\hline
&
$a^{\Lambda p}_{s}$ & $a^{\Lambda p}_{t}$ &
$a^{\Lambda n}_{s}$ & $a^{\Lambda n}_{t}$ &
$\chi^2(\Lambda p)$ & $\chi^2(\Sigma N)$ & $\chi^2({\rm total})$ \\
\hline
\hline
 NLO13(500) &-2.604 &-1.647 & -3.267 &-1.561 & 4.47 &12.13 & 16.60  \\
 NLO13(550) &-2.586 &-1.551 & -3.291 &-1.469 & 3.46 &12.03 & 15.49  \\
 NLO13(600) &-2.588 &-1.573 & -3.291 &-1.487 & 3.43 &12.38 & 15.81  \\
 NLO13(650) &-2.592 &-1.538 & -3.271 &-1.452 & 3.70 &12.57 & 16.27  \\
\hline
\hline
 NLO19(500) &-2.649 &-1.580 & -3.202 &-1.467 & 3.51 &14.69 & 18.20 \\
 NLO19(550) &-2.640 &-1.524 & -3.205 &-1.407 & 3.23 &14.19 & 17.42 \\
 NLO19(600) &-2.632 &-1.473 & -3.227 &-1.362 & 3.45 &12.68 & 16.13 \\
 NLO19(650) &-2.620 &-1.464 & -3.225 &-1.365 & 3.28 &12.76 & 16.04  \\
\hline
\end{tabular}
\renewcommand{\arraystretch}{1.0}
\end{table*}

We determine the CSB LECs from perturbative calculations of the CSB 
contribution to the $^4_\Lambda$H-$^4_\Lambda$He splittings for 
the three scenarios CSB1-3. 
Table~\ref{tab:ynprop} provides a comparison of the results for the 
different scenarios with those of the initial (NLO13 and NLO19)
$YN$ potentials, for a regulator with cutoff 
$\La = 600$~MeV, cf. Ref.~\cite{Haidenbauer:2013oca} for details. 
The total $\chi^2$ for the NLO13 and NLO19 potentials is from 
a global fit to $36$ $\La N$ and $\Si N$
data points \cite{Haidenbauer:2019boi} while the $\chi^2$ for $\La p$
includes $12$ data points \cite{SechiZorn:1969hk,Alexander:1969cx}.
In case of the scenarios CSB1 and CSB3, the CSB contributions 
are just added to the NLO13 and NLO19 potentials as published 
in Refs.~\cite{Haidenbauer:2013oca} and \cite{Haidenbauer:2019boi},
respectively.
However, for CSB2 the required CSB changes the overall $YN$ results 
considerably and the total $\chi^2$ increases to values around
$40-50$. Here we had to re-fit the charge-symmetry conserving part
in order to achieve a description of the $YN$ data that is comparable
to those in the other scenarios, cf. the $\chi^2$ values in 
Table~\ref{tab:ynprop}. After that for CSB2 the scattering length of the 
singlet state produced by the charge-symmetric part 
amounts to $a_{s}= -3.3$ fm, as compared to $-2.9$~fm for the
other two scenarios. 
We want to emphasize that the predicted binding energies of the hypertriton 
remain practically unchanged for the different considered scenarios 
for CSB. The variations are in the order of at most $30$~keV, and thus remain 
well within the experimental uncertainty. This is expected for a $T=0$ $\Lambda np$ 
state where the $\Lambda p$ and $\Lambda n$ interactions are basically averaged. 

Results for the principal scenario CSB1
are summarized in Table~\ref{tab:A1}, based on NLO13 and NLO19 and for 
the standard range of cutoffs $\Lambda =500-650$~MeV 
\cite{Haidenbauer:2013oca,Haidenbauer:2019boi}. 
Finally, the actual values for the short range CSB 
counter terms for that principal scenario are listed in Table~\ref{LECs}.
As one can see, those LECs are indeed much
smaller than the ones of the regular (charge-symmetry conserving) contact terms 
(cf. Table~3 in \cite{Haidenbauer:2013oca}) and in line with the expectations 
from power counting, see Sect.~\ref{sec:csbeft}. 

\begin{table*}
\nprounddigits{3}
\caption{The CSB LECs for the
$^1S_0$ (s) and $^3S_1$ (t) partial waves, cf.~Eq.~(\ref{VCSB}), 
fixed from the present experimental splittings 
$\Delta E(0^+) = 233$~keV and $\Delta E(1^+) = -83$~keV
(CSB1). 
}
\label{LECs} 
\begin{center}
\renewcommand{\arraystretch}{1.4}
\begin{tabular} {|l|rr|rr|}
\hline
$\Lambda$ &\multicolumn{2}{|c|}{NLO13}  & \multicolumn{2}{|c|}{NLO19} \\
\hline
\hline
&  $C_s^{CSB}$[MeV$^{-2}$] &  $C_t^{CSB}$[MeV$^{-2}$]
&  $C_s^{CSB}$[MeV$^{-2}$] &  $C_t^{CSB}$[MeV$^{-2}$] \\
\hline
 500      &   \numprint{4.691413e-3}  &  \numprint{-9.294219e-4}   &  \numprint{ 5.589635e-3} &  \numprint{-9.505238e-4}  \\
 550      &   \numprint{6.724052e-3} &   \numprint{-8.625089e-4}   &   \numprint{6.863206e-3} &  \numprint{-1.260004e-3} \\
 600      &   \numprint{9.959831e-3}  & \numprint{-9.869969e-4}    &   \numprint{9.217360e-3}  & \numprint{-1.304647e-3} \\
 650      &   \numprint{1.500219e-2}  &   \numprint{-1.142164e-3}  &   \numprint{1.240454e-2}  & \numprint{-1.394845e-3} \\
\hline
\end{tabular}
\renewcommand{\arraystretch}{1.0}
\end{center}
\end{table*} 

Let us start with the comparison of the various scenarios 
based on the calculations for the cutoff $600$~MeV, 
listed in Table~\ref{tab:ynprop}, and focus first on 
the $\Lambda n$ scattering lengths, 
Obviously there is a sizable splitting between the 
$\Lambda n$ and $\Lambda p$ results, 
depending on the CSB scenario for $A=4$ hypernuclei. 
In particular, for CSB2 the $\La n$ interaction becomes 
significantly less attractive as compared to $\Lambda p$ and, in the 
triplet state, also as compared to the case without CSB forces. 
In the other two scenarios, the singlet interaction in $\Lambda n$ 
is more attractive than that in $\Lambda p$. Furthermore, there are 
noticeably smaller changes for the triplet $\La n$ scattering length in 
those two scenarios. In particular, for CSB1 the values for 
$\Lambda n$ and $\Lambda p$ are fairly close to that without
CSB. 

Table~\ref{tab:ynprop} also provides the results of the full 
(non-perturbative) 
calculation of the CSB splittings of the $0^+$ and $1^+$ states 
for $A=4$ hypernuclei for all three 
CSB scenarios. In addition, the predictions for the original $YN$ potentials, 
without any explicit CSB force, and for the case where only the 
one-boson-exchange CSB contributions (CSB-OBE)
($\La-\Si^0$ mixing, $\eta-\pi^0$ mixing, $K^\pm/K^0$ exchange)
are added.
For CSB1 and CSB3, the CSB of the separation energy agrees within 
experimental uncertainties with the values mentioned above. For CSB2, 
there are some deviations to the 
pre-2014 situation. Given that this is an outdated scenario anyway and 
that CSB2 required a complete refit of the $YN$ interaction, we refrained 
from further improving the description of CSB.
The obtained splittings without CSB contact terms confirm the conclusion 
from earlier studies~\cite{Nogga:2001ef,Haidenbauer:2007ra,Nogga:2013pwa} 
that the standard mechanisms can only explain a very small fraction of 
the experimentally found CSB in $A=4$ hypernuclei. In particular,
because of cancellations between the OBE contributions, once 
$\eta-\pi^0$ mixing is treated properly \cite{Coon:1979si},
the overall results do not really improve when including those. 
In addition, the large variation between the NLO13 and NLO19 results 
is a clear signal for the missing CSB contact terms. 

Now we analyze in more detail the results for scenario CSB1, 
the one which is in line with 
the present experimental situation. Corresponding
results are summarized in Table~\ref{tab:A1}.
There is a clear and universal trend for a sizable splitting between 
the $\La p$ and $\La n$ scattering length in the singlet state, once
we impose the reproduction of $\Delta E(0^+)$ and $\Delta E(1^+)$.
The splitting in the triplet state is much smaller and actually
goes into the opposite direction. In particular, for reproducing the 
experimentally observed CSB splitting in the $A=4$ hypernuclei,
in the $^1S_0$ state the $\La n$ interaction is required to be more
attractive than for $\La p$, whereas for $^3S_1$ 
the $\La n$ interaction is slightly less attractive than that for $\La p$. 

With regard to the $\Lambda n$ scattering lengths the results for
the singlet channel are quite robust. The predictions are in the 
narrow range of $-3.2$ to $-3.3$~fm and practically independent on 
the cutoff and whether NLO13 or NLO19 is used. 
There is more variation in case of the triplet state which, however, 
is simply a reflection of the situation
observed already in the calculation without CSB forces.
One very interesting aspect is that, adding the CSB interaction 
to our NLO potentials established in 
Refs.~\cite{Haidenbauer:2013oca,Haidenbauer:2019boi},
improves also the overall description of the $\Lambda p$ data as 
quantified by the $\chi^2$ value -- without any refit,
see Table~\ref{tab:ynprop}. 
It is due to the noticeable reduction of the strength of the
$\Lambda p$ interaction in the singlet channel by the 
needed CSB force, cf. the pertinent scattering lengths in the
table. In fact, one could interpret this as sign for a consistency
of the available $\Lambda p$ data with the present values of
the CSB level splittings in the $A=4$ hypernuclei.
In this context we want to mention that a recent measurement of the 
$\Lambda p$ momentum correlation function in $pp$ collisions at $13$~TeV \cite{Acharya:2021fdf} likewise indicates that a slightly 
less attractive $\Lambda p$ interaction is favored by the data. 

Finally, note that $\Delta a_{1S0}^{CSB} \equiv a_{\La p} - a_{\La n}$
is $\approx 0.62\pm 0.08$~fm for the $^1S_0$ partial wave, which is 
comparable to but noticeably smaller than the CSB effects 
in the 
$pp$ and $nn$ scattering lengths where it amounts to
$\Delta a^{CSB} = a_{pp} - a_{nn} = 1.5\pm 0.5$~fm \cite{Miller:2006tv}. 
On the other hand, in case of the triplet state, the 
prediction is with  
$\Delta a_{3S1}^{CSB} \approx -0.10\pm 0.02$~fm significantly smaller and of opposite sign.

\subsection{Relation of CSB in the separation energies to the expectation values}

\begin{table}
\renewcommand{\arraystretch}{1.4}
\caption{\label{tab:h3lamhe4lamh4lamcsb} 
$^3_\Lambda$H, $^4_\Lambda$He and $^4_\Lambda$H
separation energies for scenario CSB1 for NLO13 
and NLO19 for various cutoffs in combination with SMS N$^4$LO+ (450) \cite{Reinert:2018ip} and 
 different different cutoffs for NLO13 and/or NLO19. Energies are in MeV. 
}
\begin{tabular}{l|r|rr|rr}
\hline
      interaction &  $E_\Lambda(^3_\Lambda{\rm H})$ &  \multicolumn{2}{c|}{$E_\Lambda(^4_\Lambda{\rm He})$ } &  \multicolumn{2}{c}{$E_\Lambda(^4_\Lambda{\rm H})$ }\\
                     &     &   $J^\pi=0^+$ & $J^\pi=1^+$ & $J^\pi=0^+$ & $J^\pi=1^+$ \\                                      
\hline
NLO13(500)           & 0.14 &  1.82 &  0.76 &  1.56 & 0.82  \\
NLO13(550)           & 0.10 &  1.62 &  0.56 &  1.36 & 0.61   \\
NLO13(600)           & 0.09 &  1.59 &  0.55 &  1.34 & 0.60   \\
NLO13(650)           & 0.09 &  1.61 &  0.59 &  1.36 & 0.64   \\
\hline
NLO19(500)           & 0.10 &  1.77 &  1.19 &  1.52 & 1.27   \\
NLO19(550)           & 0.10 &  1.67 &  1.21 &  1.42 & 1.28   \\
NLO19(600)           & 0.09 &  1.58 &  1.03 &  1.34 & 1.09   \\
NLO19(650)           & 0.10 &  1.65 &  0.89 &  1.40 & 0.96   \\
\hline 
Expt.                & $0.13(5)$ \cite{Juric:1973zq} 
      & 2.39(3)\cite{Yamamoto:2015avw} & 0.98(3)\cite{Yamamoto:2015avw} 
      & 2.16(8)\cite{Schulz:2016dt} & 1.07(8) \cite{Schulz:2016dt} \\
      \hline
\end{tabular}
\renewcommand{\arraystretch}{1.0}
\end{table}

This section is a short summary of the relation between the expectation values obtained 
and the CSB in the separation energies of $^4_\Lambda$He-$^4_\Lambda$H  for CSB1. We start 
with a summary of the $A=3$ and $A=4$ results for the separation energies in 
Table~\ref{tab:h3lamhe4lamh4lamcsb}. These results are qualitatively quite 
similar to the ones of Table~\ref{tab:h3lamhe4lamh4lam}. We will discuss the 
CSB contributions for $A=4$ in more detail below. Here, we just remark 
that we observed small changes of the separation energy of $^3_\Lambda$H of 
the order of 3~keV since we take isospin $T=1$ and $T=2$ components into account.
The effect of CSB is calculated here perturbatively based on the wave function 
for $^4_\Lambda$He for the orginal interaction.  
The results are given as differences of expectation values 
\begin{equation}
\langle H \rangle_{{\rm CSB}}   \equiv \langle H \rangle_{^4_\Lambda {\rm He}} - \langle H \rangle_{^4_\Lambda {\rm H}} \ .
\end{equation}
For the separation energies, one therefore finds 
\begin{eqnarray}
\Delta E_\Lambda & = &  
E_\Lambda(^4_\Lambda {\rm He}) - E_\Lambda(^4_\Lambda {\rm H})  \nonumber \\
& = & 
E(^3{\rm He}) - E(^3{\rm H}) - \left( \, E(^4_\Lambda{\rm He}) -
E(^4_\Lambda{\rm H}) \,  \right) \ \ ,
\end{eqnarray}
where then the last contribution is approximated by the expectation values. 
We separate 
\begin{eqnarray}
 & & E(^4_\Lambda{\rm He}) - E(^4_\Lambda{\rm H}) \approx \langle H \rangle_{{\rm CSB}} \nonumber \\
 & & \qquad = \langle T \rangle_{{\rm CSB}}  + 
\langle V_{NN} \rangle_{{\rm CSB}} + \langle V_{YN} \rangle_{{\rm CSB}} \ \ .
\end{eqnarray}
It turns out that $E(^3{\rm He}) - E(^3{\rm H})$ is similar 
to $\langle V_{NN} \rangle_{{\rm CSB}}$. Therefore, the contributions largely cancel. For example, one finds for the SMS $NN$ interaction 
at order N$^4$LO+ \cite{Reinert:2018ip} with cutoff $450$~MeV 
$E(^3{\rm He}) - E(^3{\rm H}) = 751$~keV. Based on the scenario CSB1 
for NLO19 (600),  $\langle V_{NN} \rangle_{{\rm CSB}} = 740$~keV.
Thereby, $E(^3{\rm He}) - E(^3{\rm H})$ contains approximately $10$~keV 
due to the mass difference of proton and neutron. The remaining $741$~keV
are mostly due to the point proton Coulomb interaction but also 
include CSB of the $NN$ interaction.  
We note that this combined effect in most cases contribute positively 
to the CSB whereas the Coulomb interaction alone is expected to give a negative contribution \cite{Bodmer:1985km,Nogga:2001ef}. 

\begin{table}
\renewcommand{\arraystretch}{1.4}
\caption{\label{tab:csbpert0} Perturbative estimate of different 
contributions to the CSB of $^4_\Lambda$He  and $^4_\Lambda$H for 
the $0^+$ state based on $^4_\Lambda$He wave functions for scenario CSB1.
The SMS N$^4$LO+ (450) $NN$ interaction \cite{Reinert:2018ip} 
was used in all cases. The contributions of the kinetic energy $\langle T \rangle_{{\rm CSB}}$, the $YN$ interaction $\langle V_{YN} \rangle_{{\rm CSB}}$ and the contribution of the nuclear core   $V_{NN}^{\rm CSB} = \langle V_{NN} \rangle_{{\rm CSB}}- E(^3{\rm He}) + E(^3{\rm H})$ are separated 
and combined to the total CSB $\Delta E_\Lambda^{pert}$. The direct 
comparison of separation energies for full calculations of 
$^4_\Lambda$He and $^4_\Lambda$H, $\Delta E_\Lambda$, is also given. 
All energies are in keV. }
\begin{center}
\begin{tabular}{l|rrr|rr}
\hline
      interaction &  $\langle T \rangle_{{\rm CSB}}$ 
                  & $\langle V_{YN} \rangle_{{\rm CSB}}$
                  &  $V_{NN}^{\rm CSB}$ 
                  & $\Delta E_\Lambda^{pert}$ 
                  & $\Delta E_\Lambda$ \\
\hline
NLO13(500)       &  44 &  200 &   16 &  261 &   265 \\
NLO13(550)       &  46 &  191 &   20 &  257 &   261 \\
NLO13(600)       &  44 &  187 &   20 &  252 &   256 \\
NLO13(650)       &  38 &  189 &   18 &  245 &   249 \\
\hline
NLO19(500)       &  14 &  224 &    5 &  243 &   249 \\
NLO19(550)       &  14 &  226 &    7 &  247 &   252 \\
NLO19(600)       &  22 &  204 &   12 &  238 &   243 \\
NLO19(650)       &  26 &  207 &   12 &  245 &   250 \\
\hline
\end{tabular}
\end{center}
\renewcommand{\arraystretch}{1.0}
\end{table}

\begin{table}
\renewcommand{\arraystretch}{1.4}
\caption{\label{tab:csbpert1} Perturbative estimate of different 
contributions to the CSB of $^4_\Lambda$He  and $^4_\Lambda$H for 
the $1^+$ state based on $^4_\Lambda$He wave functions for scenario CSB1. Same interactions and notations as in Table~\ref{tab:csbpert0}. }
\begin{center}
\begin{tabular}{l|rrr|rr}
\hline
      interaction &  $\langle T \rangle_{{\rm CSB}}$ 
                  & $\langle V_{YN} \rangle_{{\rm CSB}}$
                  &  $V_{NN}^{\rm CSB}$ 
                  & $\Delta E_\Lambda^{pert}$ 
                  & $\Delta E_\Lambda$ \\
\hline
NLO13(500)       &   5 &  -90 &   15 &  -71 &   -66 \\
NLO13(550)       &   5 &  -86 &   18 &  -63 &   -56 \\
NLO13(600)       &   4 &  -83 &   19 &  -59 &   -53 \\
NLO13(650)       &   3 &  -80 &   17 &  -59 &   -55 \\
\hline
NLO19(500)       &   1 &  -84 &    3 &  -80 &   -75 \\
NLO19(550)       &   2 &  -81 &    2 &  -77 &   -72 \\
NLO19(600)       &   4 &  -82 &    6 &  -71 &   -67 \\
NLO19(650)       &   4 &  -79 &    9 &  -66 &   -69 \\
\hline
\end{tabular}
\end{center}
\renewcommand{\arraystretch}{1.0}
\end{table}

In Tables~\ref{tab:csbpert0} and \ref{tab:csbpert1}, we summarize the 
different contributions to 
CSB for the scenario CSB1 for different $YN$ potentials. It can be seen that 
the contribution of the kinetic energy (mostly due to the mass difference 
within the $\Sigma$ multiplet) is strongly dependent on the chosen interaction. 
After properly including the CSB LECs, the $YN$ potential provides the 
by far largest contribution to the CSB. The total CSB is by construction 
fairly independent of the $YN$ interaction. The comparison of the 
perturbative estimate to the direct result for the CSB $\Delta E_\Lambda$
shows that both calculations agree well with each other. 
We note that this is also so because 
we chose $^4_\Lambda$He wave functions for the evaluation of the expectation 
values. Results 
for $^4_\Lambda$H reproduce the full calculation with  slightly lower accuracy. 

As already seen in Table~\ref{tab:A1},
also the predictions for the $\Lambda p$ and $\Lambda n$ scattering 
lengths are largely independent of the interaction. The latter property is 
not trivial and suggests that the CSB of the scattering lengths can be indeed
determined using $A=4$ data.

%%%%%%%%%%%%%%%%%%%%%%%%%%%%%%%%%%%%%%%%%%%%%%%%%%%%%%%%%

\subsection{The prescription of CSB employed by Gazda and Gal}

In the calculations by Gazda and Gal \cite{Gazda:2015qyt,Gazda:2016ir}, 
a global prescription for the CSB potential was 
employed, suggested in Ref.~\cite{Gal:2015hh}:
\begin{equation}
\langle N\La |V^{CSB}_{\La N} |N\La\rangle = -0.0297 \tau_N \frac{1}{\sqrt{3}}\langle N\Si |V|N\La\rangle \ .
\label{Gal}
\end{equation}
Obviously here the CSB contribution to the $\La N$ 
interaction is equated with the full $\La N\to \Si N$ transition potential, appropriately scaled with the $\La-\Si^0$ mixing matrix element.
Indeed, this is appropriate for isovector mesons whose direct 
contribution to the $\La N$ potential is forbidden when isospin
is conserved, but becomes non-zero via $\Lambda-\Sigma^0$ mixing. 
For example, in the Nijmegen $YN$ potentials such as NSC97 \cite{Rijken:1999fc}
this aspect is implemented and $\La-\Si^0$ mixing 
is taken into account for all isovector mesons 
($\pi$, $\rho$, $a_0$, $a_2$) that contribute to the 
$\La N \to \Si N$ transition potential. 
However, with such a global prescription CSB contributions are
also attributed to $K$ and/or $K^*$ exchange - though  
strange mesons contribute anyway directly to the 
charge-symmetry conserving part of the $\La N$ potential. 
If there is CSB from say $K$ exchange it should arise 
directly from the $\La N\to \La N$ potential, 
cf. Sect.~\ref{sec:yninter}. 

Apart from that, the prescription Eq.~(\ref{Gal}) dismisses other 
CSB contributions not related to $\La-\Si^0$ 
mixing, e.g. the ones from $\eta-\pi^0$, $\rho^0-\omega$, 
etc. mixing. According to the literature, 
$\rho^0-\omega$ mixing provides the main contribution to 
the CSB observed between the $pp$ and $nn$ 
systems \cite{Miller:2006tv}. Though pertinent studies for 
$^{\,4}_\La \rm He$ and $^{\,4}_\La \rm H$ are 
inconclusive \cite{Coon:1979si,Coon:1998jd}, they certainly reveal 
a strong model dependence. This is a clear signal 
that corresponding contact terms representing short-range 
physics should be and have to be included. 
In fact, also in case of Gazda and Gal, the result for the 
CSB splitting of the $0^+$ state based on the LO interaction 
exhibits a large cutoff dependence~\cite{Gazda:2016ir}.
For the $1^+$ state, the energy levels themselves show a 
sizable cutoff dependence, while the CSB splitting itself
appears to be too large as compared to the present
experimental value. 
In order to quantify the uncertainty of the prescription 
in Eq.~(\ref{Gal}) by ourselves, we have performed calculations for 
the $0^+$ state using the LO, NLO13 and NLO19 interactions. 
We reproduce the CSB of 
Ref.~\cite{Gazda:2016ir} for LO fairly well ($36$-$309$~keV). 
Using the same prescription, we find $461$-$2266$~keV for NLO13
and $86$-$458$~keV for NLO19. Evidently, the results at NLO  
are likewise strongly cutoff dependent and there are also 
strong variations between the two variants NLO13 and NLO19,
which otherwise yield practically identical results for $\La N$ 
and $\Si N$ scattering. 
The discussion in Section~4 of Ref.~\cite{Gazda:2016ir}
suggests that the cutoff dependence and the interplay between 
contact terms and
(unregularized) pion exchange was indeed a concern for the
authors and an attempt was made to stabilize the outcome
within an ad hoc procedure. However, adding genuine CSB
contact terms as done in the present work is the appropriate 
remedy -- and anyway required in a consistent application of chiral EFT.
Finally, we note that this again stresses that the strength 
of the $\Lambda N$-$\Sigma N$ transition potential 
is not observable and intimately correlated with consistently 
defined three-baryon interactions \cite{Haidenbauer:2019boi,Wirth:2016iwn}.

%%%%%%%%%%%%%%%%%%%%%%%%%%%%%%%%%%%%%%%%%%%%%%%%%%%%%%%
\section{Conclusions}
In the present work, we have studied effects from 
CSB in the $YN$ interaction. 
Specifically, 
we have utilized the experimentally known difference of 
the $\Lambda$ separation energies 
in the mirror nuclei ${^4_\Lambda \rm He}$ and ${^4_\Lambda \rm  H}$ 
to constrain the $\Lambda$-neutron interaction.
For that purpose, we derived the contributions of the 
leading CSB interaction within chiral effective field theory and 
added them to our NLO chiral hyperon-nucleon interactions
\cite{Haidenbauer:2013oca,Haidenbauer:2019boi}.
CSB contributions arise from a non-zero $\La\La\pi$
coupling constant which is estimated from 
$\Lambda-\Sigma^0$ mixing, the mass difference between
$K^{\pm}$ and $K^0$, and from two contact terms that
represent short-ranged CSB forces. 
In the actual calculation, the two arising CSB low-energy constants are 
fixed by considering the known differences in the energy levels
of the $0^+$ and $1^+$ states of the aforementioned
$A=4$ hypernuclei. Then, 
by construction, the resulting interaction describes 
all low energy hyperon-nucleon scattering data, the 
hypertriton and the CSB in
${^4_\Lambda \rm He}$ and ${^4_\Lambda \rm  H}$ accurately. 

It turned out that 
the reproduction of the presently established splittings of
$\Delta E(0^+) = 233\pm 92$~keV and 
$\Delta E(1^+) = -83\pm 94$~keV 
requires a sizable difference between the strength of the
$\La p$ and $\La n$ interactions in the $^1S_0$ state,
whereas the modifications in the $^3S_1$ partial wave 
are much smaller. The effects go also in
opposite directions, i.e. while for $^1S_0$ 
the $\La p$ interaction is found to be 
noticeably less attractive than $\La n$, 
in case of $^3S_1$ it is slightly more attractive. 
In terms of the pertinent scattering lengths we
predict for $\Delta a^{CSB} = a_{\La p} - a_{\La n}$
a value of $0.62\pm 0.08$~fm for the $^1S_0$ partial wave
and $-0.10\pm 0.02$~fm for $^3S_1$.

The required CSB implies a significantly stronger $\La n$ 
interaction in the $^1S_0$ partial wave and the pertinent
scattering length of our NLO potentials 
\cite{Haidenbauer:2013oca,Haidenbauer:2019boi}
increases from $-2.9$ fm to around $a^{\Lambda n}_s=-3.2$ fm.
Therefore, it is worthwhile to explore in how far this
has consequences for the possible existence of $\La nn$ resonances 
\cite{Garcilazo:2014lva,Gal:2014efa,Hiyama:2014cua,Richard:2014pwa,Ando:2015fsa,Afnan:2015ahc,Kamada:2016ozg,Gibson:2019occ,Gibson:2020npg,Schafer:2020vzl,Schafer:2020jnb}.
It will be also interesting to utilize the CSB forces established 
in the present work in calculations of $p$-shell hypernuclei.
Indeed, there are several experimentally established mirror
hypernuclei in the $p$-shell region with $A=7-10$ \cite{Botta:2017eg,Davis:2005npa,Botta:2019has} which have
been already studied within phenomenological approaches in
the past \cite{Hiyama:2009ki,Hiyama:2012sq,Hiyama:2013owa}.
However, now those systems can be also explored by more
systematic microscopic approaches such as {\it ab-initio} 
calculations based on the no-core shell model 
\cite{Wirth:2017bpw,Wirth:2019cpp,Le:2020zdu}.

Interestingly, adding the CSB interaction to our NLO potentials 
established in 
Refs.~\cite{Haidenbauer:2013oca,Haidenbauer:2019boi},
improves slightly the overall description of the
$YN$ data as quantified by the $\chi^2$ value -- without
any refit. This could be interpreted as sign for a consistency
of the available $YN$ data with the present values of
the CSB level splittings in the $A=4$ hypernuclei.
In any case, with regard to the latter new but still preliminary 
results by FINUDA for the ground-state binding energy of 
$^{\,4}_\La \rm H$ have 
been reported already \cite{Botta:2019has} and 
further measurements of the splitting of the $0^+$ and $1^+$ 
states of $^{\,4}_\La \rm H$ are planned at 
J-PARC~\cite{Akazawa:2016,Yamamoto:2019dcf}.

%%%%%%%%%%%%%%%%%%%%%%%%%%%%%%%%%%%%%%%%%%%%%%%%%%%%%%%
\vskip 0.4cm
\begin{acknowledgements}
We acknowledge stimulating discussions with Avraham Gal and
Josef Pochodzalla.
This work is supported in part by the DFG and the NSFC through
funds provided to the Sino-German CRC 110 ``Symmetries and
the Emergence of Structure in QCD'' (DFG grant. no. TRR~110)
and the VolkswagenStiftung (grant no. 93562).
The work of UGM was supported in part by The Chinese Academy
of Sciences (CAS) President's International Fellowship Initiative (PIFI)
(grant no.~2018DM0034). We also acknowledge support of the THEIA net-working activity 
of the Strong 2020 Project. The numerical calculations were performed on JURECA
and the JURECA-Booster of the J\"ulich Supercomputing Centre, J\"ulich, Germany.
\end{acknowledgements}

\bibliographystyle{unsrturl}

\bibliography{bib/hyp-exp.bib,bib/ncsm.bib,bib/hyp-theory.bib,bib/nn-interactions.bib,bib/yn-interactions.bib,bib/faddeev-yakubovsky.bib,bib/srg.bib}

\end{document}